\documentclass[10pt, conference]{IEEEtran}
\IEEEoverridecommandlockouts
\usepackage[english]{babel}
\usepackage{blindtext}
\usepackage{multirow}
\usepackage{booktabs} % <-- line 2
\usepackage{cite}
\usepackage{amsmath,amssymb,amsfonts}
\usepackage{algorithmic}
\usepackage{graphicx}
\usepackage{textcomp}
\usepackage{xcolor}
\def\BibTeX{{\rm B\kern-.05em{\sc i\kern-.025em b}\kern-.08em
    T\kern-.1667em\lower.7ex\hbox{E}\kern-.125emX}}
    
\usepackage{tikz}

\usepackage{times}  
\usepackage[hidelinks]{hyperref}
\usepackage{graphics}
\usepackage{flushend}

\usepackage{epsfig}
\usepackage{color}
\usepackage{caption} 

\usepackage{sidecap}
\usepackage{subcaption}

\newcounter{todocounter}

\newcommand{\name}{Hercules\xspace}

\newcommand{\ignore}[1]{}

\newcommand{\hide}[1]{}

\newtheorem{theorem}{Theorem}[section]

\def\shownotes{1}   % set 1 for version with author notes
\ifnum\shownotes=1
\definecolor{mygreen}{rgb}{0,0.5,0}
\definecolor{myred}{rgb}{0.5,0,0}

\newcommand{\vect}[1]{\langle #1 \rangle}
\newcommand{\myfair}[0]{HRF}
\begin{document}
\title{\name: Heterogeneous Requirements Congestion Control Protocol}
%\title{Hercules: Congestion Control for Heterogeneous Applications}
%\author{Paper \# 1571019203}
%\author{\IEEEauthorblockN{Neta Rozen-Schiff, Amit Navon, Itzcak Pechtalt and Leon Bruckman}}

\author{\IEEEauthorblockN{Neta Rozen-Schiff}
%\IEEEauthorblockA{\textit{Network Time Foundation} \\
%}
\and
\IEEEauthorblockN{Amit Navon}
%\IEEEauthorblockA{\textit{Check Point Software} \\
%}
\and
\IEEEauthorblockN{Itzcak Pechtalt}
%\IEEEauthorblockA{\textit{Unifabrix}\\
%}
\and
\IEEEauthorblockN{Leon Bruckman}
%\IEEEauthorblockA{\textit{Huawei Technologies} \\
%}
}

\maketitle
\begin{abstract}
%Today's networks are required to serve diverse applications with diverse requirements.
Future network services present a significant challenge for network providers due to high number and high variety of co-existing requirements. 
%Today's networks are struggling to scale and satisfy the high number and high variety of co-existing network requirements.  
Despite many advancements in network architectures and management schemes, congested network links continue to constrain the Quality of Service (QoS) for critical applications like tele-surgery and autonomous driving.
A prominent, complimentary approach consists of congestion control (CC) protocols which regulate bandwidth at the endpoints before network congestion occurs. However, existing CC protocols, including recent ones, are primarily designed to handle small numbers of requirement classes, highlighting the need for a more granular and flexible congestion control solution.

In this paper we introduce \name, a novel CC protocol designed to handle heterogeneous requirements. \name is based on an online learning approach and has the capability to support any combination of requirements within an unbounded and continuous requirements space. 
We have implemented \name as a QUIC module 
%(planned to be open source) 
and demonstrate, through extensive analysis and real-world experiments, that \name can achieve up to 3.5-fold improvement in QoS compared to state-of-the-art CC protocols.
%can outperform state-of-the-art CC protocols QoS by a factor of $3.5$.
%that in some cases, \name can improve QoS by up to $250\%$, outperforming state-of-the-art CC protocols.

\end{abstract}

\begin{IEEEkeywords}
Congestion Control, QoS requirements, Online learning
\end{IEEEkeywords}

\section{Introduction}\label{sec:intro}

Over the years, networks have improved significantly in terms of speed and reliability. However, even modern network architectures, such as $5$G/$6$G networks, are facing congestion due to the escalating bandwidth requirements \cite{6G_congestion_1,6G_congestion_2,congestion_6G}.
These requirements are driven by the rising popularity of real-time streaming applications like Augmented Reality (AR), Virtual Reality (VR), and the proliferation of IoT devices \cite{IoT_CC1}. 

\begin{figure}%[t]
    \centering
    \includegraphics[width=0.8\linewidth]{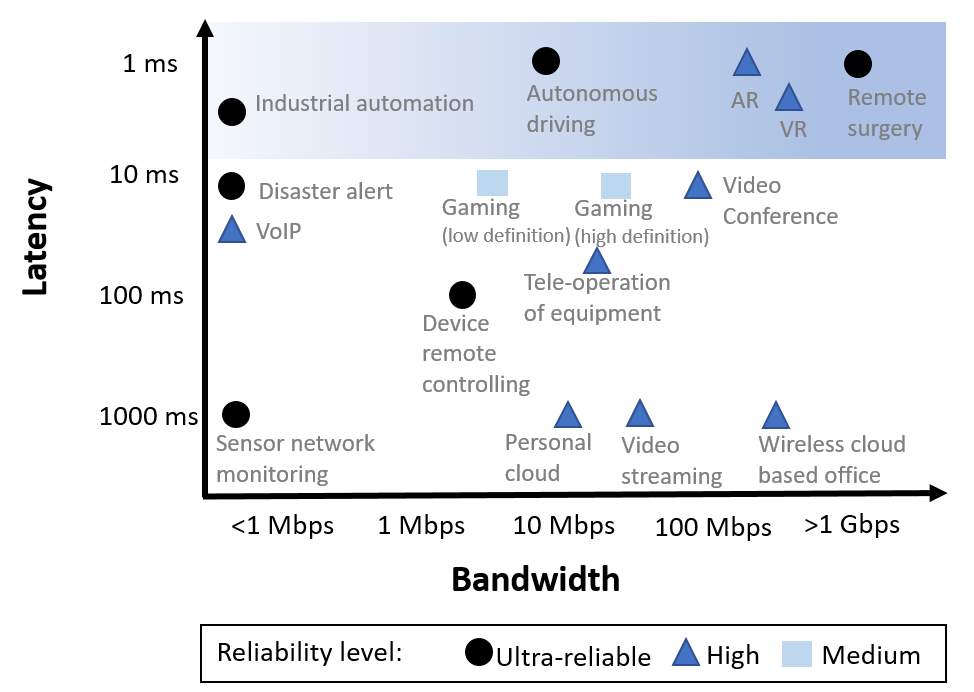}
    \caption{Delay, bandwidth and reliability requirements for different applications}
     \label{fig:motivation}
     %\vspace{-0.7cm}
\end{figure}

Congested networks are, by definition, limited in their ability to meet all service requirements. Among common factors like bandwidth, reliability, and latency, the highest heterogeneity is typically observed in bandwidth, ranging from $10$ Kbps (e.g., for industrial automation) to $100$ Mbps (e.g., for $3$D video streaming)  \cite{BW_vs_delay1,BW_vs_delay_sensitive}. Fig.~\ref{fig:motivation} provides a summary of various application requirements regarding bandwidth, latency, and reliability. Applications that experience the most significant Quality of Service (QoS) degradation in congested networks are those requiring high bandwidth and low latency, such as AR, VR, and remote surgery (grouped in a blue rectangle).

In order to satisfy high QoS requirements, service providers 
use several approaches. These include bringing services closer to consumers through edge computing, proxies etc., enhancing in-network routing and scheduling policies based on service prioritization.
Another important approach involves regulating the sending rate at the endpoints using congestion control (CC) protocols. This helps to reduce excessive network load, which could otherwise lead to delays and packet loss.

CC protocols were originally designed to achieve ''fair share`` bandwidth allocation, 
treating all applications uniformly
%, regardless of their requirements 
\cite{BBR,cubic,PCC, vivace,COPA}. 
Clearly, such protocols are sub-optimal in heterogeneous settings involving co-existing network services with different requirements.
%\neta{However, they fail to satisfy high bandwidth requirement applications \cite{PRISM}.}
%\subsection{Heterogeneous CC protocols}
%Moreover, they were unable to meet heterogeneous requirements \emph{between} connections of the same application. For instance, an AR application might comprise control, audio, video, and tactile connections. Among these, control, audio, and tactile connections are typically more time-sensitive and crucial than higher-bandwidth video connections, which can tolerate resolution reductions without significant Quality of Experience (QoE) impact. The goal, is therefore to enable heterogeneous connections to coexists while delivering the highest possible Quality of Service (QoS).
%As a result, recent studies have delved into unfair bandwidth allocations, suggesting that ''the era of friendly TCPs has been broken down`` \cite{FB_TCP}.

As a result, many studies have delved into unfair bandwidth allocations.
However, the existing ''unfair'' CC protocols have one or more of the following limitations: (i) assume perfect knowledge of the network in order  to define a goal weight of each connection \cite{CC_shares,FairCloud_CC} (ii) treat connections according to a limited number of priority classes \cite{Cubic_Ledbat_Overview, ledbat_plus, proteus}, or (iii) specifically tailored for a single type of application, such as IoT \cite{IoT_CC1, IoT_CC2, IoT_CC3} and video streaming \cite{vivace, PRISM}.

%based on prioritization \cite{cc_overview_2020},

%The ''unfair'' CC protocols can be categorized as either: (i) weight-based  \cite{CC_shares,FairCloud_CC, economical_CC} which assume perfect knowledge of the network, (ii) priority-based protocols, that support two different application requirements  \cite{Cubic_Ledbat_Overview, LEDBAT_rfc,ledbat_plus, proteus}, or (iii) protocols specifically tailored for a single type of application, such as IoT \cite{IoT_CC1, IoT_CC2, IoT_CC3}.

%Another effort focused on a bandwidth allocation mechanism for unbou\cite{AQUA} (instead of CC protocol), but as such, it was limited to applications with the same source and destination.

%In this work we introduce \emph{\name}, a generic CC protocol designed to regulate bandwidth in unfair manner which supports co-existences of unbounded levels of heterogeneous requirements. 
%which maximize the number of  an unlimited levels of heterogeneous requirement within the network.
%, varies by their endpoints and requirements. 
%that provides fairness w.r.t application requirements. To this end, we define \emph{Heterogeneous Requirements based Fairness (\myfair)} which is suited for today's network applications. \myfair~prioritizes connections based on their relative requirement satisfaction thereby providing more attention for connections with low bandwidth requirements on the expense of high bandwidth connections.

In this work we introduce \emph{\name}, a CC protocol designed to achieve an (unfair) bandwidth allocation that satisfies any combination of heterogeneous requirements.

Our main contributions include:
\begin{itemize}
     %\item \noindent{We establish the desired fairness for heterogeneous applications}.
     %\item \noindent{\bf present a theoretical analysis of Hercules CC} and 
     \item Leveraging insights from online-learning, we design \name, a provable CC protocol that ensure fairness based on heterogeneous requirements.
    \item We have implemented \name as a CC module for QUIC and made it publicly available \cite{Hercules_code}.
    \item We evaluate \name through both theoretical analysis and real-world experiments.
    \end{itemize}

  %   \item We share AQUA's code\footnote{\url{https://github.com/itzcak/AQUA}} as a patch that can be easily deployed on top of QUIC.

\section{Preliminaries and Motivation}\label{sec:background_goals}
%\neta{Congestion control protocols run independently on each endpoint and regulate the sending rate of each connection base on its shared bottleneck.} 
Online learning has emerged as an effective approach for bandwidth regulation, as seen in recent CC protocols like Vivace \cite{PCC,vivace} and it derivatives \cite{MPCC,proteus}.
These protocols continually collect feedback from the network, including  loss ratio and Round-Trip-Time (RTT) data derived from ACK packets. Using this information, they estimate the current congestion level and dynamically adjust the sending rate of endpoints.
Each sender autonomously operates within the protocol, making decisions about adjusting the sending rate based on a utility function, similar to a player in game theory.

The primary objectives of these protocols are twofold: (i) ensuring fairness among connections and (ii) maximizing the utilization of the network resources.

The utilization objective typically aims to achieve Pareto efficiency, representing a network state where a sender cannot increase its rate without reducing others'. 

Fairness, on the other hand, is more challenging to define in scenarios with heterogeneous requirements, as discussed next.
%\neta{Fairness, on the other hand, is more challenging to define. The naive ''fair-share'' approach, where bandwidth is equally divided among connections, can result in suboptimal outcomes, especially in scenarios with heterogeneous requirements. %Therefore, the naive approach was extended as discussed next.
%} 
%This objective alone is easy to obtain by just having each connection increasing its sending rate until experiencing congestion, but clearly this won't yield a fair allocation. On the other hand, setting all sending rates to zero can be considered fair but it  clearly doesn't utilize the network.

%However meeting these two objectives is not an easy task. 

\subsection{Existing Fairness Definitions}  
Online learning CC protocols, such as Vivace \cite{PCC,vivace}, typically employ the {\bf ''fair-share''} principle as their fundamental fairness principle. This principle divides bandwidth equally among competing connections. However, this definition is suboptimal for heterogeneous connections \cite{PRISM}.

%\neta{The fundamental fairness principle applied in online learning CC protocols (like Vivace \cite{PCC,vivace}) is known as {\bf ''fair-share''}, which divides bandwidth equally among competing connections at the endpoint. However, it is unable to address heterogeneous connections \cite{PRISM}.} 

\hide{
{\bf Max-min fairness (MMF)} \cite{max_min1, max_min2} extends the "fair-share" principle by aiming to maximize the bandwidth allocated to the most poorly treated connection \cite{max_min2}.  
However, in certain scenarios, MMF solution is not Pareto efficient \cite{LMMF}.
%As a result, MMF becomes inefficient 
For example, in networks with heterogeneous requirements, MMF may impose 
the rate limit of one connection on all others. 
%a connection with limited resource impose the other connections the same limited sending rate 
%For example, as 
%On top of the fact that MMF does not achieve Pareto efficiency\footnote{Pareto efficiency is a point where a sender can not increase its rate, unless a lower rate sender's rate is decreased}, 
Moreover, MMF allocation does not always exist \cite{max_min_may_not_exists}.
%, and even when it does, it does not capture the senders' heterogeneous requirements.
}
Lexicographic max-min fairness, also known as {\bf Max-Min Fairness (MMF)},  extends the "fair-share" principle by aiming to maximize the bandwidth allocated to the most poorly treated connections \cite{max_min2}.
Consider the connection rates as a vector, $X=\vect{x_0, ..., x_n}$, whose elements are ordered from low to high, MMF is defined as the lexicographically maximal vector that can be achieved\cite{LMMF}. In other words, MMF vector %doesn't consist of just the max min rate ($\argmax{x_0}$) but also the
consists of the
maximums of any 
%other 
rate ($x_i$) given the values of the lower rates ($x_0,...,x_{i-1}$).  
%LMMF can be seen as searching for a lexicographically maximal vector (of rates) in the space of the feasible vectors with components rearranged in non-decreasing order \cite{LMMF}. 
%As a result, LMMF maximizes not only the minimum connection but also the others. 
%bandwidth assigned to the worse connection but also maximizes the second-to-worst connection, third and so on 

\begin{figure}%[htbp]
%\hspace{0.1\columnwidth}
 \centering
%\begin{subfigure}{0.33\linewidth}
%  \centering
%  \includegraphics[width=\linewidth]{figs/MMF.png}
  %\vspace{-0.5cm}
%  \caption{MMF}
%  \label{fig:MMF}
%\end{subfigure}%
\begin{subfigure}{0.5\linewidth}
  \centering
  \includegraphics[width=\linewidth]{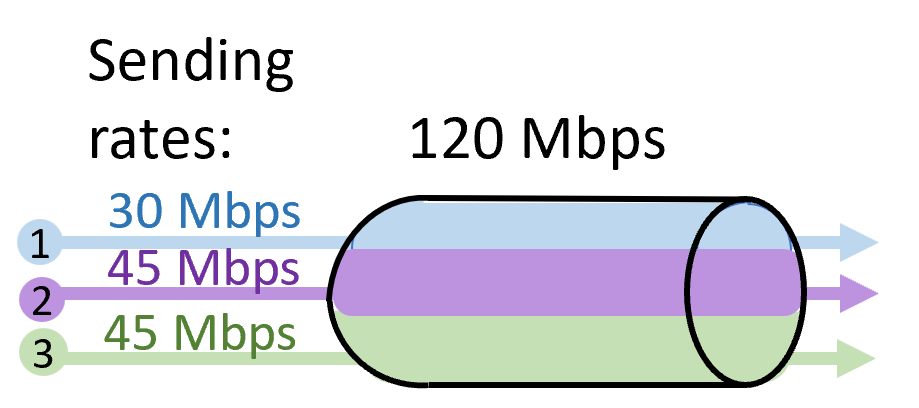}
  %\vspace{-0.5cm}
  \caption{MMF}
  \label{fig:LMMF}
  %\vspace{-0.3cm}
\end{subfigure}%
\begin{subfigure}{0.5\linewidth}
  \centering
  \includegraphics[width=\linewidth]{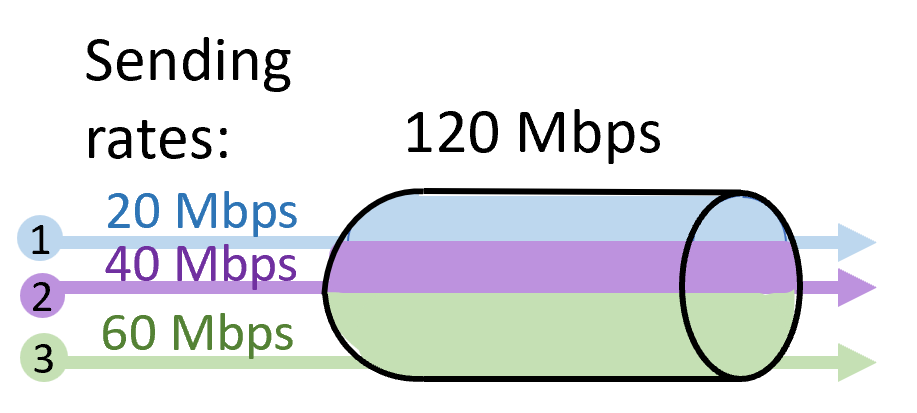}
  %\vspace{-0.5cm}
  \caption{Optimal}
  \label{fig:optimal}
  %\vspace{-0.3cm}
\end{subfigure}%

%\vspace{-0.4cm}
\begin{subfigure}{\linewidth}
  \centering
  \includegraphics[width=\linewidth]{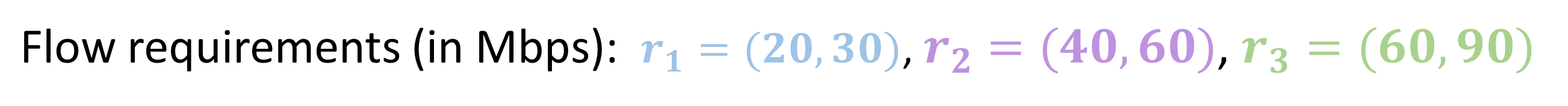}
  %\vspace{-0.5cm}
\end{subfigure}%
\caption{Different fairness objectives} 
\label{fig:fairness}
\vspace{-0.5cm}
\end{figure}

MMF achieves Pareto efficiency \cite{MPCC, LMMF} and exists as long as the number of connections is finite \cite{LMMF_exists}.
Recent CC protocols (such as MPCC \cite{MPCC}) were developed with the goal of converging to MMF. However, MMF
%Although MMF mitigates the limitations of MMF, it 
can lead to suboptimal outcomes in scenarios involving three or more requirement levels, as demonstrated next.
\subsection{Heterogeneous Requirements and Fairness}\label{sec:Heterogeneous Requirements and Fairness}
%However, today's network applications have different network requirements. For example, HD video streaming with $20$Mbps connections, 3D video with $100$Mbps connections and tactile channels with $500$Kbps connections. 
%In many cases, these application can operate in a wide range of bandwidths, for example video streaming may support several picture resolutions and or frame rates. 
Given $n$ connections with heterogeneous requirements, we denote $(a_i, b_i)$ for each connection $i$, where $a_i$ and $b_i$ are the minimum and maximum bandwidth requirements respectively. The minimum requirement is set as the threshold below which the application will experience significant degradation in service quality. The maximum requirement is defined as the rate at which an application will not experience significant improvement and typically will refrain from increasing the sending rate. Note that the minimum and maximum requirements can be provided by the application or inferred using traffic classification \cite{traffic_classification1,traffic_classification2}.
%Flow Feature-Based Network Traffic Classification Using Machine Learning

For example, consider three connections, with requirements of $r_1=(20, 30)$ Mbps, $r_2=(40, 60)$ Mbps and $r_3=(60, 90)$ Mbps and a bottleneck capacity of 120Mbps. Fig.~\ref{fig:fairness} illustrates the sending rates for each connection obtained using MMF along with the optimal sending rate. 
%In MMF (Fig.\ref{fig:MMF}) a fair-share approach is enforced, resulting in all connections sending at an equal rate of 30Mbps, which leads to under utilization of the network. Furthermore, the minimum requirements of two connections are not met. On the other hand. 
MMF (Fig.\ref{fig:LMMF}) fully utilizes the network in terms of sending rates, but still fails to meet the minimum requirements of the third connection (r3). However, an optimal solution (Fig.\ref{fig:optimal}) can utilize the network and satisfy the requirements of all connections.

Hence, to tackle the challenge of heterogeneous requirements among connections, a new fairness approach is required. This approach should take into account the specific requirements of each connection and ensure that, in congested scenarios, at least the minimum requirement of each connection is met, if possible. 

\section{\name Design Overview}\label{sec:Hercules_design}
We present \name, a CC protocol that extends the online learning approach for heterogeneous connections.
%(described in Section~\ref{sec:online_learning}) 
%in order to guarantee convergence to \neta{heterogeneous connection fairness}. 
%which considers the heterogeneous requirements of connections. 
We begin by defining heterogeneous fairness, then we describe \name's utility function. Finally we introduce the rate control module which uses the utility function to regulate sending rates.

\ignore{
\begin{table}[t]\centering
%\begin{tabular}{|p{5cm}||p{3.2cm}|p{3.2cm}|p{3cm}|p{1.5cm}|}
\begin{tabular}{|p{1cm}p{6.5cm}|}
 \hline
 %Video Quality & \multicolumn{2}{|c|}{\text{Bandwidth Requirements}} \\  
 %\cline{2-3}
 {\bf Notation}  & {\bf Description}  \\
 \hline\hline
 $a_{i}$ & Minimum requirement of connection $i$\\
 %\hline
 $b_{i}$ & Maximum requirement of connection $i$\\
 %\hline
$x_{i}$ & Sending rate of connection $i$\\
 %\hline
$\overline{x_{i}}$ & Relative reward of connection $i$  w.r.t its requirements.\\
 %\hline
$H(\overline{x_i})$ & Requirement penalty of connection $i$ w.r.t the weighted rate.\\
 %\hline
$L_i$ & Loss experienced by connection $i$ \\
%\hline
$RTT_i$ & The RTT experienced by connection $i$ \\
\hline
\end{tabular}\caption{Notation Table}
\vspace{-0.6cm}
\label{table:notation table}
\end{table}
}

\begin{figure}%[t]
    \centering
    \includegraphics[width=0.55\linewidth]{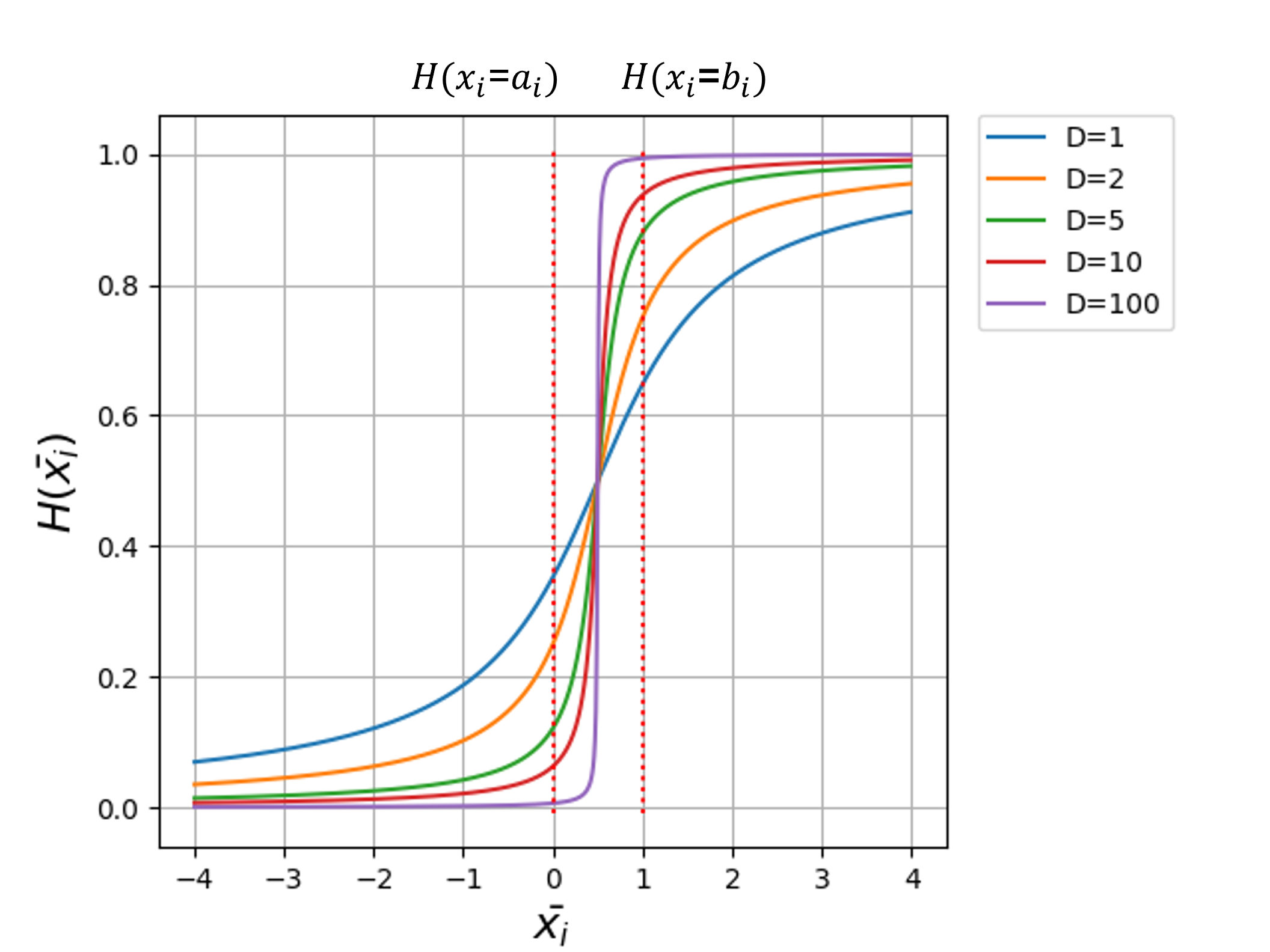}
    \caption{Different requirement penalties as functions of normalized rates, for different values of D. The normalized rate values $0$ and $1$ correspond to the minimum and maximum requirements respectively, and are denoted in red dashed lines.}
     \label{fig:arctan}
\end{figure}

\subsection{Heterogeneous Requirements Fairness}\label{sec:HRF}
%Existing fairness objectives, such as LMMF, were defined only in terms of connection rates regardless connection requirements.
In order to maximize the satisfaction of connection requirements we
extend MMF by 
considering the sending rates when normalized by their requirements. For each connection $i$ with sending rate $x_i$ we define its normalized rate by 
$\overline{x_i} = \frac{x_{i}-a_{i}}{b_{i}-a_{i}}$, 
where $(a_i, b_i)$ are the minimum and maximum requirements of connection $i$ respectively.
Note that when the sending rate of connection $i$ is within its minimum and maximum requirements (i.e., $x_i \in [a_i,b_i]$), its normalized rate is within $0$ and $1$ (i.e., $\overline{x_i} \in [0,1]$).

Our Heterogeneous Requirements Fairness metric (\myfair) is defined as the lexicographical max min vector of the normalized rates, ordered from low to high (i.e., $\vect{\overline{x_1},...,\overline{x_n}}$ where $\overline{x_{i-1}} \leq \overline{x_i}$ for every $i$). 
For example, considering the three connections case in Section~\ref{sec:Heterogeneous Requirements and Fairness} (Fig.~\ref{fig:optimal}), the optimal sending rates of $\vect{x_1,x_2,x_3}=\vect{20,40,60}$  are obtained when the lexicographical max min of the normalized sending rates $\vect{\overline{x_1},\overline{x_2},\overline{x_3}}=\vect{0,0,0}$ is obtained.

%Note that the normalized rates used in the vector are scaled according to the requirements, which amplifies unsatisfied requirements of low rates connections compared to those with higher rates. For instance, a sending rate of $9$ Mbps for a connection  with requirement $(10,20)$ is considered much more ``critical'' than a sending rate of $99$ Mbps for a connection with requirement $(100,200)$.

%Therefore, \myfair~ ensures requirements satisfaction when possible while also minimizing unsatisfaction scaled according to requirements.
%\myfair~ is achieved where every connection $i$ can not increase his utility on expends on the connection with higher relative reward.

\subsection{\name Utility Function}\label{sec:utility_function_section}
\name leverages from the rich literature on online learning and utility functions for CC \cite{PCC,vivace,proteus}. On top of them, 
%\neta{The utility function of \name leverages from the rich literature of online learning functions \cite{PCC,vivace,proteus} by  including a }
%\name utility function shares the same congestion penalties of Proteus scavengers utility function, including loss, RTT gradient and deviation. %However, 
in order to converge to \myfair{}, \name utility function also includes a 
''\emph{requirement penalty}``, denoted as $H(\overline{x_i})$, which depends on the normalized sending rate.

Inspired by the activation functions in neural networks \cite{arctan_neu_networks}, the requirement penalty uses the $\arctan$ function for scaling and is defined as follows:
%The requirement penalty uses the $\arctan$ function for scaling similar 
%The requirement penalty adjusts all the congestion penalties on a scale of $0$ to $1$ using the $\arctan$ function.
%\footnote{$\arctan$ is also used as activation function in neural networks \cite{arctan_neu_networks}.}.
%This term increases as the sending rate exceeds the minimum requirement. 
%The resulting utility function is defined as follows for every connection $i$:
%The requirement penalty of connection $i$ is defined as follows:
\begin{align} \label{eq:H(x)}
    H(\overline{x_i}) &= \frac{\arctan(D \cdot  (\overline{x}_{i} - \frac{1}{2}))}{\pi}+\frac{1}{2} \in (0,1),
\end{align}
%where $D\geq 1$ is a coefficient which expresses the penalty magnitude for rates below or above their required interval $(a_i, b_i)$.
where $D\geq 1$ is a coefficient which scales the requirement penalty especially when sending rates get closer to the required rate bounds $(a_i, b_i)$ and beyond. 

%\neta{We analyzed the requirement penalty for different values of $D$.}
%Our findings reveal that 
Fig.~\ref{fig:arctan} demonstrates different requirement penalties as functions of normalized rates, for different values of $D$. Rates within the requirement interval correspond to normalized rates within the interval $[0,1]$ which is denoted in red dashed lines.
%We can see that

Note that higher $D$ values imply higher penalties for connections exceeding their maximum requirement and lower penalties for connections falling below their minimum requirement. However, when $D$ is excessively high, the penalty function becomes abrupt and approaches a step-like function numerically. This could potentially affect convergence.
%(for a detailed discussion on convergence, see Section \ref{sec:convergance}). In our experiments, we used $D = 2$, which yielded superior results. 
%In Section \ref{sec:convergance} we explore the impact of $D$ on convergence in real experiments and following it suggest to use $D=2$.
In Section \ref{sec:convergance} we analyze the impact of $D$ on convergence in real experiments. Following this analysis we choose $D=2$ as the suggested value.

Finally, \name utility function is defined as follows:
\begin{align}\label{eq:Hercules_utility}
    U(x_{i}) =& (x_{i})^t -  x_{i} \cdot H(\overline{x_i}) \cdot [\beta \cdot L_{i} 
     - \gamma \cdot max\{0,\frac{d(RTT_{i})}{dt}\} \nonumber \\
    & - \varphi \cdot \sigma(RTT_i)],
\end{align}

\begin{figure}[t]
    \centering
    \includegraphics[width=\linewidth]{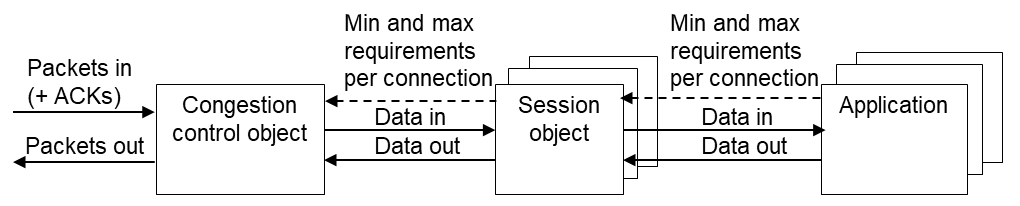}
    \caption{\name' extension over QUIC objects}
     \label{fig:Hercules_implementation}
       \vspace{-0.45cm}
\end{figure}

where $L_i$ and $RTT_i$ are the loss and RTT experience by connection $i$ respectively\footnote{The coefficient values are $t=0.9$, $\beta = 11.35$ (enables up to $5\%$ random loss rate tolerance), $\gamma = 25$ and $\varphi  = 750$).}. 
%$t=0.9$, $\beta = 11.35$, $\gamma = 25$ and $\varphi  = 750$.
%but with different values for the latency and deviation (
%$\gamma = 25$ and $\varphi  = 750$ respectively.
Note that when no latency is experienced, the utility function is much simpler: $U(x_{i}) = (x_{i})^t$. 
%Table \ref{table:notation table} summarizes \name notations.

%In \cite{Hercules_proofs} we prove Theorem \ref{theo:eq_LMMF} which states that if all senders use the \name utility function and converge to equilibrium their sending rates will be \myfair{}. 

\begin{theorem}\label{theo:eq_LMMF}
In a network with $n$ connections competing over a bottleneck, where the connections are regulated by \name per connection 
utility function (\ref{eq:Hercules_utility}) with per connection requirements $\{(a_i,b_i)\}_{i \in [n]}$, any equilibrium is \myfair{}.
\end{theorem}

Due to space constraints, we provide only a brief outline of the proof. First, we show that with the absence of congestion rates increase and therefore congestion will occur at some point in time. Next we prove that during congestion, unsatisfied connections with lower requirements exhibit a steeper utility gradient compared to satisfied connections with higher requirements. Lastly, we conclude by establishing that there is no equilibrium in which connections with lower requirements can increase their rates at the expense of connections with higher requirements.

%The proof, detailed in \cite{Hercules_proofs}, 
%\neta{Due to space constraints, we provide only a brief outline of the proof. It} begins by demonstrating that at some point in time, congestion will occur. Next, it proceeds to illustrate that during congestion, unsatisfied connections with lower requirements exhibit a steeper utility gradient compared to satisfied connections with higher requirements. The proof concludes by establishing that there is no equilibrium in which connections with lower requirements can increase their rates at the expense of connections with higher requirements.

\subsection{\name Rate-Control Module}\label{sec:rate_control}

\name adopts an online learning approach by dividing time into sequential update intervals. At the beginning of each interval, a connection can be in one of three states: (i) slow-start, (ii) probing, or (iii) moving. %\name modifies these states and their transition conditions according to the connection's specific requirements.
%For completion, the three states are defined as follows:
%The description of all three states is provided next: 

%\vspace{0.05in}\noindent{\bf Slow start,} 
%When a connection starts sending,
During the \emph{\bf slow start} phase, the initial sending rate is determined as the minimum between a predefined rate value (configured to $5$ Kbps in our evaluation) and the connection's minimum requirement. Subsequently, in each time interval, the sending rate is doubled unless either the utility function decreases or the sending rate exceeds the connection's maximum requirement. In such cases, the state transitions to probing.

%\vspace{0.05in}\noindent{\bf Probing state} 
%In this state the utility gradient is estimated and used to determine whether the connection’s rate should increase or decrease. 
In the \emph{\bf probing} state, we calculate the utility gradient using the current sending rate ($x$). This is done by comparing the utilities achieved when sending $x+\delta x$ and $x-\delta x$, where $\delta$ is set to be a fraction (e.g., $5\%$) of $x$.  The gradient determines whether the connection's rate should increase or decrease.  
%Once the direction (increase / decrease) decision is made the moving begins.
%Specifies whether the connection’s rate should increase or decrease. Given the current rate of r of a connection, the utility's gradient is computed for $r \pm v$. Once the direction decision is made, the connection switches to the moving state.

%\vspace{0.05in}\noindent{\bf Moving state} 
During \emph{\bf moving} state, the sending rate continues to change in the same direction as determined by the last probing decision until the utility decreases, indicating a change in gradient direction. At this point, the protocol transitions back to probing.

\begin{theorem}\label{theo:eq_point}
In a network with $n$ connections competing over a bottleneck, where the connections are regulated by \name per connection 
utility function (\ref{eq:Hercules_utility}) with per connection requirements $\{(a_i,b_i)\}_{i \in [n]}$, the sending rates converge to an equilibrium.
\end{theorem}

\begin{figure}%[htbp]
 \centering
%\vspace{-0.4cm}
\begin{subfigure}{0.4\linewidth}
  \centering
  \includegraphics[width=\linewidth]{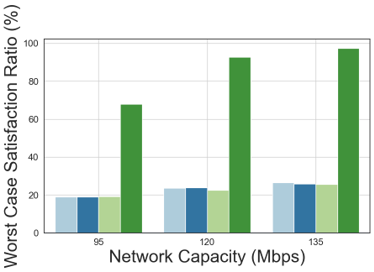}
  \vspace{-0.5cm}
  \caption{Unbounded}
  \label{fig:worse_unbounded}
\end{subfigure}%
\hspace{0.5em}% Space between image A and B
\begin{subfigure}{0.4\linewidth}
  \centering
  \includegraphics[width=\linewidth]{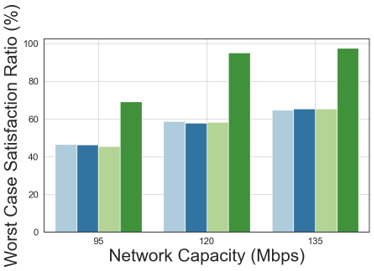}
  \vspace{-0.5cm}
  \caption{Single unbounded}
  \label{fig:worse_unspecified}
\end{subfigure}%
%\vspace{-0.4cm}
\hspace{0.5em}% Space between image A and B
\begin{subfigure}{0.16\linewidth}
  \centering
  \includegraphics[width=\linewidth]{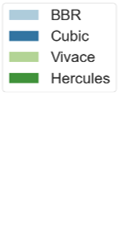}
  \vspace{-0.5cm}
  %\vspace{-0.3cm}
\end{subfigure}%
%\vspace{-0.4cm}
\caption{Satisfaction ratio for the two scenarios at different network congestion conditions}
\label{fig:worse_sat}
\vspace{-0.3cm}
\end{figure}

\begin{figure*}%[htbp]
 \centering
%\vspace{-0.4cm}
\begin{subfigure}{0.2\linewidth}
  \centering
  \includegraphics[width=\linewidth]{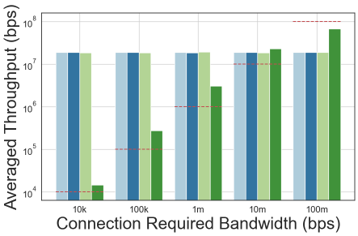}
  %\vspace{-0.5cm}
  \caption{$95$ Mbps}
  \label{fig:Lab 120 throughput_bar_prioritization}
\end{subfigure}%
\hspace{1em}% Space between image A and B
\begin{subfigure}{0.2\linewidth}
  \centering
  \includegraphics[width=\linewidth]{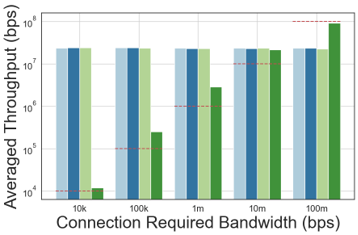}
  %\vspace{-0.5cm}
  \caption{$120$ Mbps}
  \label{fig:Lab 120 throughput_bar_prioritization}
\end{subfigure}%
\hspace{1em}% Space between image A and B
\begin{subfigure}{0.2\linewidth}
  \centering
  \includegraphics[width=\linewidth]{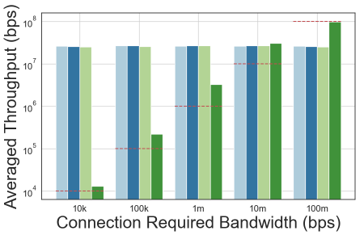}
 % \vspace{-0.5cm}
  \caption{$135$ Mbps}
  \label{fig:Lab 135 throughput_bar_prioritization}
  %\vspace{-0.3cm}
\end{subfigure}%
\hspace{1em}% Space between image A and B
\begin{subfigure}{0.1\linewidth}
  \centering
  \includegraphics[width=\linewidth,trim={0  1.25cm 0 0},clip]{figs/allocated_BW_legend.png}
%  \vspace{-0.5cm}
  %\vspace{-0.3cm}
\end{subfigure}%
%\vspace{-0.4cm}
\caption{Average throughput for all unbounded scenario at different network congestion conditions. The red dashed line denotes the connections' requirements.}
\label{fig:AWS_throughput_bar}
\vspace{-0.3cm}
\end{figure*}

Due to space constraints, we provide only a sketch of the proof.
%The proof, detailed in \cite{Hercules_proofs}, 
First, we establish that \name' utility function is concave and so the system consisting of senders using the rate-control module constitutes a concave game. 
Then we conclude the proof since a concave
game has an equilibrium point \cite{Even_dar}.

\section{Implementation}\label{sec:implementation}

\name' implementation (available online \cite{Hercules_code}) extends QUICHE, 
%\footnote{Version ID $407956543$, released on Nov the $5$th, $2021$.}, 
a widely deployed C$++$ implementation of QUIC (also used in Google Chrome). 
As illustrated Fig.~\ref{fig:Hercules_implementation}, \name' rate control logic  mainly resides in two QUICHE components: (i) the congestion control object and (ii) the session object.

\vspace{0.05in}\noindent{\bf Session object}
is responsible for managing the connection parameters.
\name extends this object by including the minimum and maximum requirements received from the application, which are then passed on to the congestion control object. Additionally, \name introduces scheduled calls at regular intervals, prompting the congestion control object to calculate and update the sending rate accordingly.

\vspace{0.05in}\noindent{\bf Congestion control object} has been extended to carry out the rate control module (as discussed in Section~\ref{sec:rate_control}) based on the minimum and maximum requirements received from the session object.
Specifically, during each scheduled call initiated by the session object,  the 
CC object measures the RTT and loss rate (derived from receiving ACKs), calculates the current state and utility function, and subsequently 
 updates the new sending rates.
%Specifically, we implemented both Hercules and Vivace \cite{vivace} utility function and states transitions. 
%Fig.~\ref{fig:Hercules_implementation} describes Hercules implementation over QUICHE. 

\begin{figure*}%[htbp]
 \centering
%\vspace{-0.4cm}
\begin{subfigure}{0.2\linewidth}
  \centering
  \includegraphics[width=\linewidth]{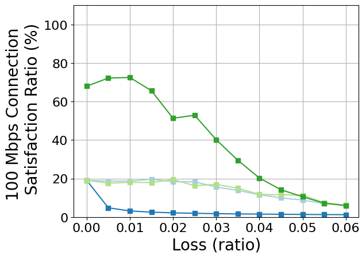}
  %\vspace{-0.5cm}
  \caption{$95$ Mbps}
  \label{fig:loss_120}
\end{subfigure}%
\hspace{1em}% Space between image A and B
\begin{subfigure}{0.2\linewidth}
  \centering
  \includegraphics[width=\linewidth]{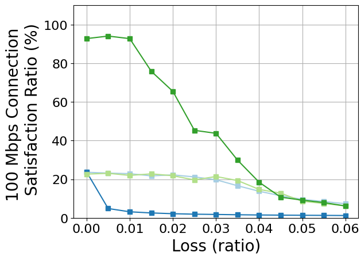}
  %\vspace{-0.5cm}
  \caption{$120$ Mbps}
  \label{fig:loss_120}
\end{subfigure}%
\hspace{1em}% Space between image A and B
\begin{subfigure}{0.2\linewidth}
  \centering
  \includegraphics[width=\linewidth]{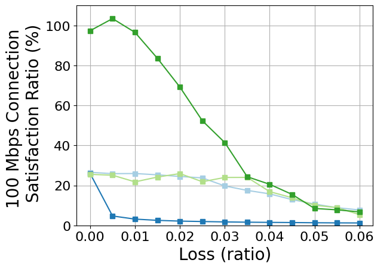}
 % \vspace{-0.5cm}
  \caption{$135$ Mbps}
  \label{fig:loss_135}
  %\vspace{-0.3cm}
\end{subfigure}%
\hspace{1em}% Space between image A and B
\begin{subfigure}{0.1\linewidth}
  \centering
  \includegraphics[width=\linewidth,trim={0 0.8cm 0 0},clip]{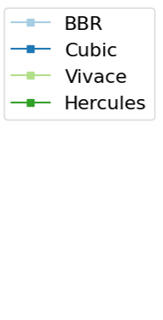}
%  \vspace{-0.5cm}
  %\vspace{-0.3cm}
\end{subfigure}%
%\vspace{-0.4cm}
\caption{Loss tolerance of the $100$ Mbps requirement connection, for the unbounded scenario}
\label{fig:AWS_loss}
\vspace{-0.3cm}
\end{figure*}
\section{Evaluation}\label{sec:evaluation}

We evaluate \name by benchmarking its performance against other state-of-the-art CC protocols, including BBR \cite{BBR}, CUBIC \cite{cubic} and Vivace \cite{vivace} across a range of network conditions. This evaluation is conducted through both laboratory simulations and real-world experiments.
 
%through lab and real-word experiments with different network conditions. 
%\subsection{Evaluation Framework} 

%\vspace{0.05in}\noindent{\bf Lab and live experiments.} 
The lab experiments consist a setup with a sender machine, a receiver machine, and a third machine that simulates various network conditions. %which run Ubuntu 20.04 and 
%connected through a third machine which emulates different network conditions.

Additionally, we conducted live experiments on the Internet, spanning both Europe and the US. In Europe, we use a receiver located in London and a sender located in Frankfurt. In the US we use two AWS receivers machines, positioned in north California and north Virginia, along with two sender machines in Ohio and Oregon, respectively.
All the machines run Ubuntu 20.04 OS.
The network capacity was changed by running traffic control utility (the \emph{tc} command) on the sender side, with the receiver reporting the total received bytes per second.

Due to space constraints, we present the live experiments results in the US, where the receiver is located in  north California and the sender is in Oregon. Similar results were obtained in the other locations in the US, Europe, and in the lab.

The evaluation results are presented next, addressing the following challenges encountered by CC protocols:
%Given the evaluation framework above, we focus on 
%three main aspects of heterogeneous requirement satisfaction: (i) guarantees in worst and average case (ii) convergence time, and (iii) fairness between Hercules and other CC protocols. 
 (i) satisfying heterogeneous requirements in both worst and average-case scenarios, (ii) assessing convergence time, and (iii) evaluating fairness compared to other CC protocols.

%\vspace{0.05in}\noindent{\bf Live experiments} conducted 
%We also ran live experiments 
%in Europe and in the US. In US we use two AWS machines - a client (located in Ohio) and a server (located in north California). In Europe  the client located in London and a server in located in Frankfurt. All the machines run Ubuntu 20.04 OS.
%The network capacity was changed by running traffic controller (TC) on the server side. In addition, for simulate random loss, we run network emulator (Netem).
%\subsection{Heterogeneous Requirements and Network Conditions}
%\label{sec:Requirements and Network Conditions}

\subsection{Satisfying Heterogeneous Requirements}\label{sec:Satisfying Heterogeneous Requirements}

\vspace{0.05in}\noindent{\bf Heterogeneous Settings.}
%\vspace{0.05in}\noindent{\bf Heterogeneous Scenarios.}
Following Section~\ref{sec:intro}, we examine coexistence of connections with heterogeneous requirements: $10$ Kbps, $100$ Kbps, $1$ Mbps, $10$ Mbps, $100$ Mbps.
The maximum requirements are set at 1.5 times higher, providing a $50\%$ margin. Due to space constraints, we concentrate on two scenarios based on the applications generating these connections:
    
%\neta{Due to lack of space, we focus on two scenarios w.r.t} 
%regards to 
%the applications generating the connections:
%, we considered two scenarios:
\begin{enumerate}
    \item All the applications have unbounded sending rate (which means they can exceed their maximum requirements).
    %\item All connections are bounded (e.g., Zoom streaming, VoIP, IoT devices).
    \item All applications except one have a bounded sending rate, while one application has an unbounded sending rate.
\end{enumerate}
Note that unbounded applications represent network services such as file downloads and database batch processing and synchronization.  In contrast, bounded applications may consist of Zoom streaming, VoIP, IoT devices etc.

\begin{figure*}[t]
 \centering
%\vspace{-0.4cm}
\begin{subfigure}{0.2\linewidth}
  \centering
  \includegraphics[width=\linewidth]{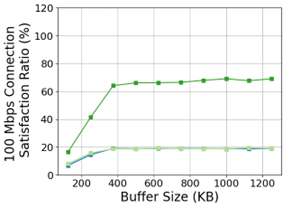}
  \vspace{-0.5cm}
  \caption{$95$ Mbps}
  \label{fig:loss_120}
\end{subfigure}%
\hspace{0.5em}% Space between image A and B
\begin{subfigure}{0.2\linewidth}
  \centering
  \includegraphics[width=\linewidth]{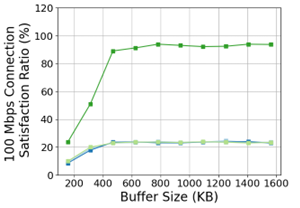}
  \vspace{-0.5cm}
  \caption{$120$ Mbps}
  \label{fig:loss_120}
\end{subfigure}%
\hspace{0.5em}% Space between image A and B
\begin{subfigure}{0.2\linewidth}
  \centering
  \includegraphics[width=\linewidth]{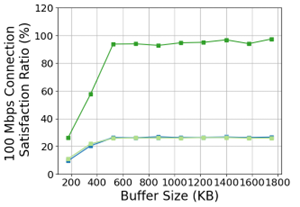}
  \vspace{-0.5cm}
  \caption{$135$ Mbps}
  \label{fig:loss_135}
  %\vspace{-0.3cm}
\end{subfigure}%
\hspace{0.5em}% Space between image A and B
\begin{subfigure}{0.1\linewidth}
  \centering
  \includegraphics[width=\linewidth,trim={0 0.55cm 0 0},clip]{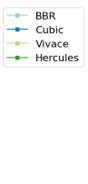}
  \vspace{-0.5cm}
  %\vspace{-0.3cm}
\end{subfigure}%
%\vspace{-0.4cm}
\caption{Bottleneck saturation with varying buffer size, for $100$ Mbps requirement connection, in the unbounded scenario}
\label{fig:AWS_buff}
\vspace{-0.3cm}
\end{figure*}

%\vspace{0.05in}\noindent{\bf Network conditions.}
We analyzed bottlenecks with $20$, $40$ and $60$ms RTT. Due to space constraints and consistent results across all bottlenecks, we focus on presenting the results for the $20$ ms bottleneck. Considering the combined minimum and maximum requirements are $111.11$ and $166.65$ Mbps respectively, 
we evaluated \name' performance  under various congestion levels:
\begin{itemize}
    \item High congestion - $95$ Mbps capacity
    \item Medium congestion - $120$ Mbps capacity
    \item Low congestion - $135$ Mbps capacity
\end{itemize}

%\vspace{0.05in}\noindent{\bf Comparison.} 
The sending rates for each connection are compared between \name to Vivace, BBR and CUBIC, across $70$ trials, each lasting $30$ seconds.
%We consider $3$ scenarios (of bounded and unbounded streams) as defined in Section~\ref{sec:Evaluation_Framework}.

We define the \emph{''satisfaction ratio``} provided by a CC protocol for a connection $i$ in a given scenario, as the ratio between the average sending rate (in that scenario) and the minimum requirement ($a_i$) of that connection. 
Moreover, we consider the worst case (i.e., guaranteed) satisfaction ratio of a protocol as the minimum satisfaction ratio across all scenarios.

\ignore{
\begin{table}%[t]\centering
%\begin{tabular}{|p{5cm}||p{3.2cm}|p{3.2cm}|p{3cm}|p{1.5cm}|}
\begin{tabular}{|p{1.3cm}|p{1.35cm} p{1.35cm} p{1.35cm} p{1.35cm}|}
 \hline
 Capacity (Mbps) & \multicolumn{4}{|c|}{\text{CC protocols Utilization ($\%$)}} \\  
 \cline{2-5}
   & BBR & CUBIC & Vivace & \name  \\
 \hline
 $95$ & 94.79 & 94.80 & 94.19 & 94.23 \\
% \hline
 $120$ & 94.79 & 94.80 & 93.49 & 93.99 \\
% \hline
 $135$ & 94.79 & 94.80 & 93.25 & 93.94 \\
 \hline
\end{tabular}\caption{Utilization ($\%$) of the tested CC protocols for every network capacity} 
\vspace{-0.3cm}
\label{table:utilization_table}
\end{table}
}

\begin{figure}[t]
%\hspace{0.1\columnwidth}
 \centering
%\vspace{-0.4cm}
\begin{subfigure}{0.33\linewidth}
  \centering
  \includegraphics[width=\linewidth]{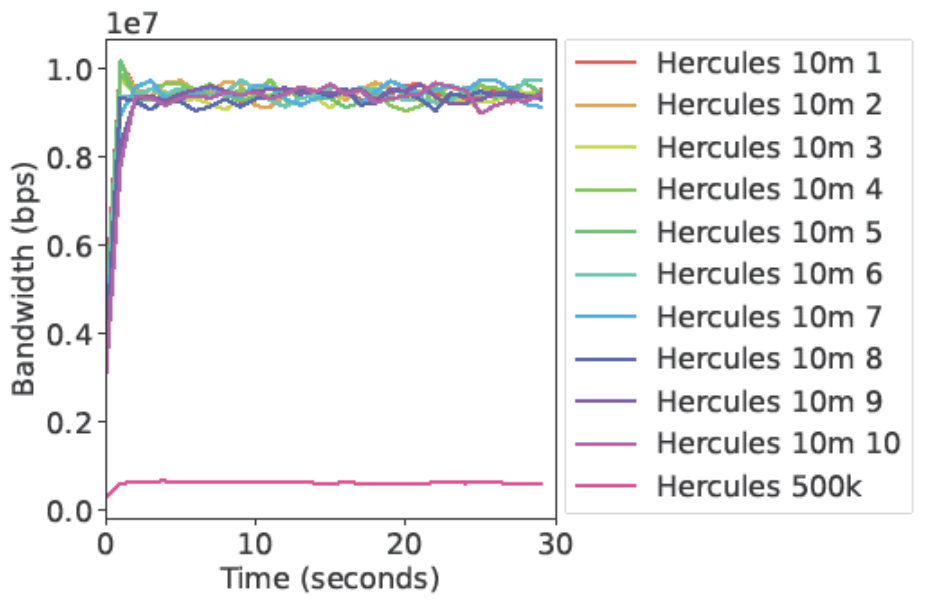}
%  \vspace{-0.5cm}
  \caption{$2$ requirements levels}
  \label{fig:static_10 hercules flows 100Mbps and scav}
%  \vspace{-0.3cm}
\end{subfigure}%
\hspace{0.5em}% Space between image A and B
%  \hspace{1em}% Space between image A and B
\begin{subfigure}{0.33\linewidth}
  \centering
  \includegraphics[width=\linewidth]{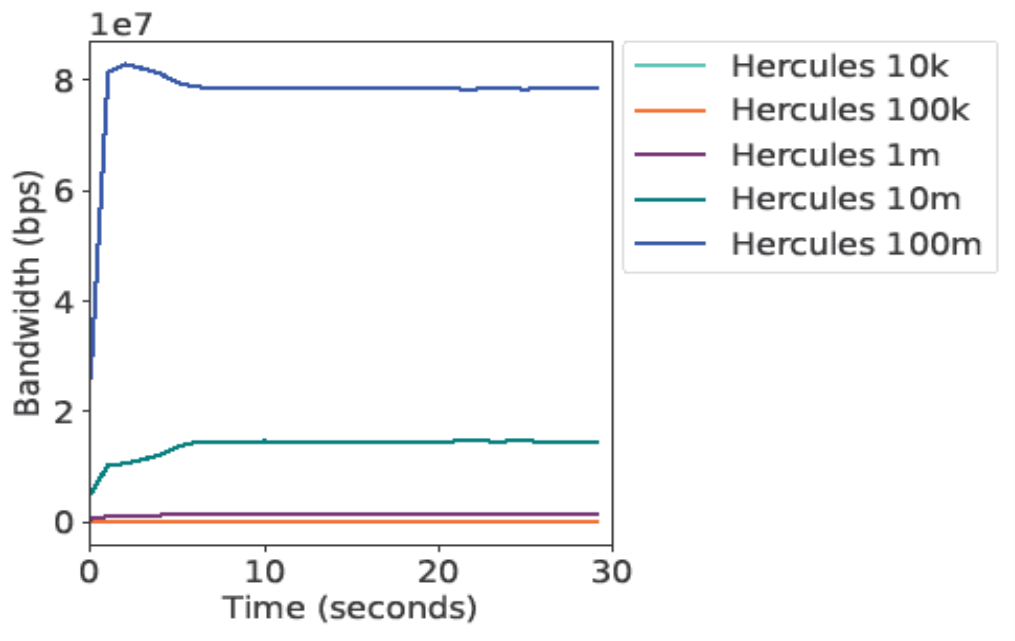}
%  \vspace{-0.5cm}
  \caption{$5$ requirements levels}
  \label{fig:static_5 hercules flows}
%  \vspace{-0.3cm}
\end{subfigure}%
\hspace{0.5em}% Space between image A and B
%\hspace{1em}% Space between image A and B
\begin{subfigure}{0.3\linewidth}
  \centering
  \includegraphics[width=\linewidth]{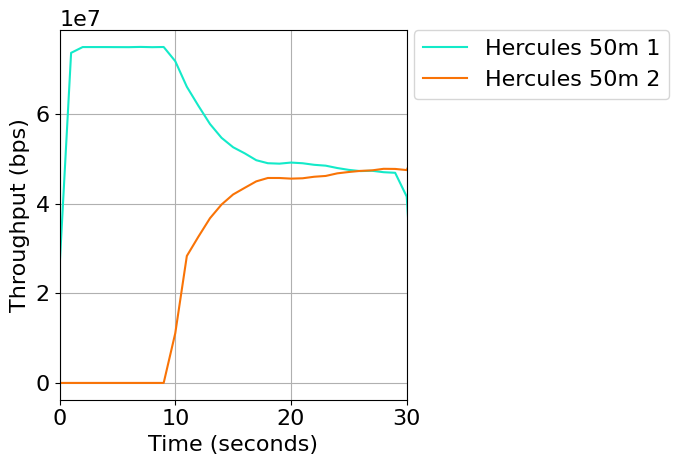}
%  \vspace{-0.5cm}
  \caption{$2$ dynamic requirements levels}
  \label{fig:dynamic_20 hercules flows}
%  \vspace{-0.3cm}
\end{subfigure}%
\caption{Throughput of \name connections}
\vspace{-0.3cm}
\label{fig:static_conv}
\end{figure}

\ignore{
\begin{figure}[t]
%\hspace{0.1\columnwidth}
 \centering
%\vspace{-0.4cm}
\begin{subfigure}{0.5\linewidth}
  \centering
  \includegraphics[width=\linewidth]{figs/Scenario7.png}
%  \vspace{-0.5cm}
  \caption{Connections with same requirements.}
  \label{fig:dynamic_10 hercules flows 100Mbps and scav}
%  \vspace{-0.3cm}
\end{subfigure}%
\hspace{1em}% Space between image A and B
\begin{subfigure}{0.4\linewidth}
  \centering
  \includegraphics[width=\linewidth]{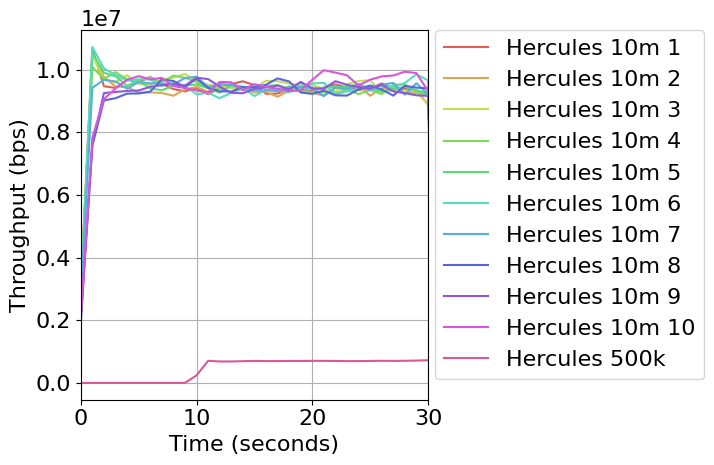}
 % \vspace{-0.5cm}
  \caption{Connections with different requirements.}
  \label{fig:dynamic_10 hercules flows 100Mbps and late scav}
\end{subfigure}%
\caption{Throughput of non simultaneous connections.}
\vspace{-0.3cm}
\label{fig:dynamic_conv}
\end{figure}}

%Fig.~\ref{fig:AWS_throughput_bar} presents the distance between allocated bandwidth and requirements, under high congestion, for unbounded connections scenario.  
%$100\%$ is denoted by an horizontal dashed red line, since it reflects the minimum value for which all the connections are satisfied.}
%\neta{It is shown that Hercules advantage over the other CC protocols is mainly achieved in the unbounded scenario.}

\vspace{0.05in}\noindent{\bf Satisfaction ratio} 
%\vspace{0.05in}\noindent{\bf Single unbounded scenario results.} 
achieved by each of the CC protocols under different network conditions is presented in Fig.~\ref{fig:worse_sat}, for both single unbounded and all unbounded connection scenarios. 
The findings reveal that across all network conditions, \name outperforms the other evaluated protocols by a factor of $1.5$ and $3.5$ in the single-unbounded and all-unbounded scenarios respectively.

%\vspace{0.05in}\noindent{\bf All unbounded scenario results.} 
Fig.\ref{fig:AWS_throughput_bar} provides a detailed analysis of the all unbounded scenario, showing the average throughput across 
%\neta{A closer look at the all unbounded scenario is provided} in Fig.\ref{fig:AWS_throughput_bar}, \neta{which present the average throughput at 
various congestion levels. 
%the poor performance of other CC protocols in the all-unbounded scenario, is due to the fact that they equally share the capacity between the connections. As a result, 
Even during low congestion ($135$ Mbps) where all connections can theoretically be satisfied, the other CC protocols fall short.
%\neta{It shows that even under low congestion (e.g, $135$ Mbps), where all the connections can be satisfied, the other CC protocols are far from optimal. 
For instance, the $100$ Mbps connection only achieves $27\%$ of its requirement due to fair-share principles, whereas with \name it achieves $97\%$ satisfaction.
In medium congestion level ($120$ Mbps), \name maintains a $92\%$ satisfaction for this connection. Furthermore, during high congestion ($95$ Mbps capacity), the optimal throughput reaches only $85\%$ satisfaction, with \name achieving $70\%$.
%Overall, \name's results are  close to optimal throughput, obtaining $97\%$ and $92\%$ worst case satisfaction for low and medium congestion ($135$ and $120$ Mbps) respectively. 
%In high congestion case ($95$ Mbps capacity), the optimal throughput is only $85\%$ satisfaction and \name obtains $70\%$.

\vspace{0.05in}\noindent{\bf Random loss tolerance.}
The satisfaction evaluation of the 100Mbps connection was extended to include scenarios with random loss. 
%occurs, we compare different CC protocols' average throughput from multiple $30$-second runs on the same bottleneck $5$ BDP for different capacities. 
%specifically, we examine the protocols stability for the $100$ Mbps satisfaction ratio across bounded and single unbounded scenarios. 
%Due to lack of space, we focus on the loss tolerance in the unbounded scenario (see ). 
Fig.~\ref{fig:AWS_loss} illustrates the satisfaction ratio achieved in the all unbounded scenario.
The results indicated that, \name outperformed all other evaluated protocols, especially for loss rate up to $4\%$.
These results are consistent with those observed in the single unbounded scenario.%, and consolidate with BBR and Vivace, for higher loss rates. %rates\footnote{The loss tolasimilarity at the point of $4.5\%$ onward is due to the  due to Hercules similar value of $\beta$ coefficient's in the utility function, in Equation (\ref{eq:Hercules_utility}). 

\vspace{0.05in}\noindent{\bf Buffer size tolerance.} 
The robustness of satisfaction ratio was analyzed across a broad range of buffer sizes $0.5$-$5$ BDP, under various network conditions (Fig.~\ref{fig:AWS_buff}). 
%Due to space limitation, we focus on the unbounded connections scenario, and 
%. For presentation clarity, Fig.~\ref{fig:buffer} 
%specifically, on the $100$Mbps connection, which can achieve $97\%$ satisfaction by Hercules (as discussed above).
%which is satisfied only by adding AQUA under these network conditions (as presented in 
%Fig.~\ref{fig:AWS_buff} presents the satisfaction ratio of the $100$ Mbps connection, for the varying buffer sizes. 
%It shows that all protocols are stable from $2$-$5$ BDP, where
The results show that in all unbounded scenario, \name outperform all protocols across the tested buffer sizes, even with shallow buffers. Similar results were obtained in a single unbounded scenario. 

\vspace{0.05in}\noindent{\bf Utilization} is calculated as the total sending rate across all connections divided by the network capacity. 
We found that BBR and CUBIC exhibit similar median utilization values across all network conditions, ranging between $94.79\%$ (BBR) and $94.8\%$ (CUBIC). \name and Vivace achieved slightly lower median utilization of $94.23\%$ and $94.19\%$ respectively.
%of the all unbounded scenario (for each protocol and network condition) \neta{ranged between $93.94\%$ in the }
%are summarized in Table~\ref{table:utilization_table}. 
%It shows that \name utilize the network to high degree (about $94\%$) similar to other state-of-the-art protocols. 
The results for the single unbounded scenario were similar. 

\newcommand{\test}[0]{test}
\newcommand{\tests}[0]{tests}

\begin{figure}[t]
    \centering
    \begin{subfigure}{0.5\linewidth}
    \includegraphics[width=\linewidth]{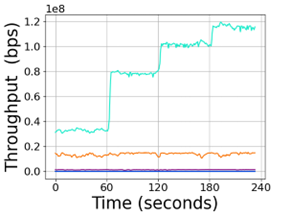}
     \end{subfigure}%
     \begin{subfigure}{0.158\linewidth}
  \centering
  \includegraphics[width=\linewidth]{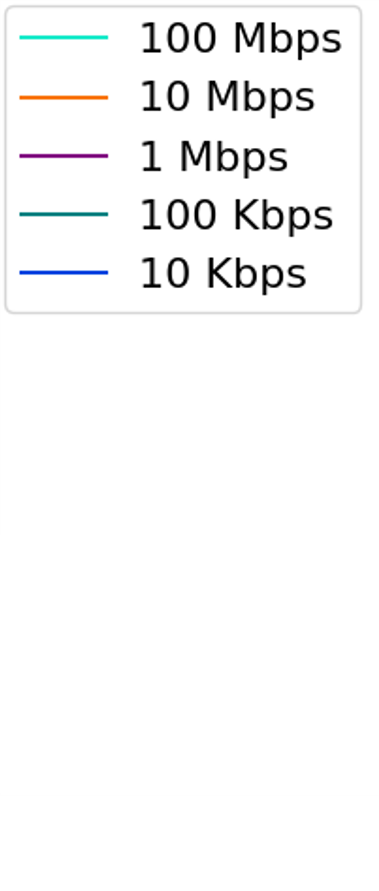}
  \vspace{-0.5cm}
  %\vspace{-0.3cm}
\end{subfigure}%
\caption{\name sending rates when network capacity changes every $60$ seconds between $45$, $95$, $120$ and $135$ Mbps}
     \label{fig:over_time}
 \vspace{-0.3cm}
\end{figure}

\subsection{Convergence Time}\label{sec:convergance}
Next we examine the convergence time of \name in three different tests: (i) \emph{static \tests{}}, involving simultaneous connections, (ii) \emph{dynamic \tests{}}, with non simultaneous connections and (iii) \emph{dynamic network \tests{}}, where network conditions (specifically 
 capacity) change during the testing phase.
 
\begin{figure}%[htbp]
 \centering
%\vspace{-0.4cm}
  \vspace{-0.5cm}
\begin{subfigure}{0.4\linewidth}
  \centering
  \includegraphics[width=\linewidth]{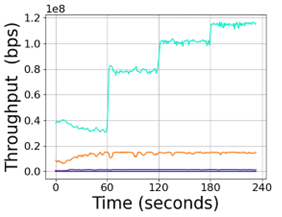}
  \vspace{-0.5cm}
  \caption{$D=1$}
  \label{fig:loss_120}
\end{subfigure}%
\begin{subfigure}{0.4\linewidth}
  \centering
  \includegraphics[width=\linewidth]{figs/var_test_D=2.png}
  \vspace{-0.5cm}
  \caption{$D=2$}
  \label{fig:loss_120}
\end{subfigure}%
\begin{subfigure}{0.15\linewidth}
  \centering
  \includegraphics[width=\linewidth]{figs/var_legend.png}
  \vspace{-0.5cm}
  %\vspace{-0.3cm}
\end{subfigure}%

\begin{subfigure}{0.4\linewidth}
  \centering
  \includegraphics[width=\linewidth]{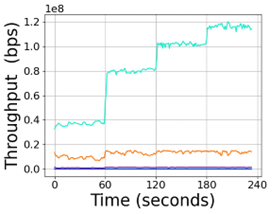}
  \vspace{-0.5cm}
  \caption{$D=5$}
  \label{fig:loss_135}
  %\vspace{-0.3cm}
\end{subfigure}%
\begin{subfigure}{0.4\linewidth}
  \centering
  \includegraphics[width=\linewidth]{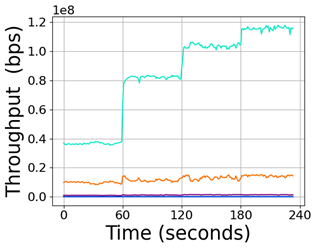}
  \vspace{-0.5cm}
  \caption{$D=10$}
  \label{fig:loss_120}
\end{subfigure}%
\begin{subfigure}{0.15\linewidth}
  \centering
  \includegraphics[width=\linewidth]{figs/var_legend.png}
  \vspace{-0.5cm}
  %\vspace{-0.3cm}
\end{subfigure}%

%\vspace{-0.4cm}

\begin{subfigure}{0.4\linewidth}
  \centering
  \vspace{-0.4cm}
  \includegraphics[width=\linewidth]{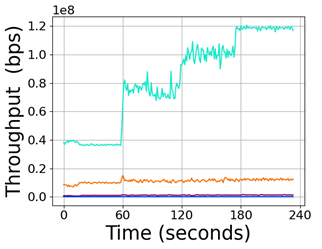}
  \vspace{-0.5cm}
  \caption{$D=100$}
  \label{fig:loss_120}
\end{subfigure}%
\begin{subfigure}{0.4\linewidth}
  \centering
  \includegraphics[width=\linewidth]{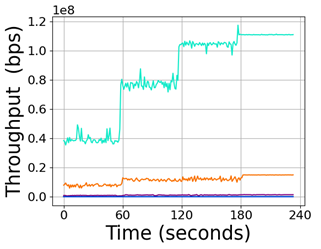}
  \vspace{-0.5cm}
  \caption{$D=1000$}
  \label{fig:loss_135}
  %\vspace{-0.3cm}
\end{subfigure}%
\begin{subfigure}{0.15\linewidth}
  \centering
  \includegraphics[width=\linewidth]{figs/var_legend.png}
  \vspace{-0.5cm}
  %\vspace{-0.3cm}
\end{subfigure}%
%\vspace{-0.4cm}
\caption{Requirement penalty (D) impact on convergence}
\label{fig:var_test_D}
\vspace{-0.3cm}
\end{figure}

\vspace{0.05in}\noindent{\bf Static \tests{}} include two scenarios using $95$ Mbps bottleneck. The first scenario considers  $10$ \name connections, each requiring $10$ Mbps and a low requirement $0.5$ Mbps connection. The results (in Fig.~\ref{fig:static_10 hercules flows 100Mbps and scav}) show
 fast convergence with all \name connections approaching their requirement.
 %, where the low requirement connection is closer ($96\%$ satisfaction).  
Similar results are obtained when the bottleneck includes $5$ different requirement levels, $10$ Kbps up to $100$ Mbps (Fig.\ref{fig:static_5 hercules flows}).
%presented in Section~\ref{sec:Satisfying Heterogeneous Requirements}.  presents the convergence in this case.}

\vspace{0.05in}\noindent{\bf Dynamic arrival \test{}.} 
In Fig.~\ref{fig:dynamic_20 hercules flows}  we display \name' sending rates for two connections, both requiring $50$ Mbps, competing over a $95$ Mbps bottleneck, where one of them arrives $10$ seconds after the first one. 
Initially, the sending rate of the first connection dominates the network utilization. Approximately $5$ seconds after the second connection arrives, their sending rates converge to \myfair{} (which in this case equivalent to fair share). 
Comparable outcomes were observed for connections with higher requirement levels.

\vspace{0.05in}\noindent{\bf Dynamic network \test{}.} 
We further stress \name stability and robustness by examining its reaction to network changes. We consider connections with $5$ requirement levels (ranging from $10$ Kbps to $100$ Mbps), with network capacity changes every $60$ seconds. In addition to the previously analyzed capacity levels ($95$ Mbps, $120$ Mbps and $135$ Mbps), we also introduce $45$ Mbps capacity (very high congestion).
The results in Fig.~\ref{fig:over_time} show that sending rates converge quickly, within a few seconds after each capacity change.

\vspace{0.05in}\noindent{\bf Dynamic network \test{} with different $D$ values.} 
The dynamic network \test{} was repeated to evaluate the influence of parameter $D$ in \name utility function (Equation (\ref{eq:H(x)})) on convergence.
%, aligning with the \neta{theoretical analysis} in Section~\ref{sec:utility_function_section}. 
The tests results, presented in Fig.~\ref{fig:var_test_D}
consider several $D$ values in the range $[1, 1000]$.
%presents convergence time plots for various  parameter values ($D\in\{1,2,5,10,15,100,1000\}$). 
We observed that for $D=1$, the initial convergence is slower compared to higher $D$ values, while extremely high values of $D$ (e.g., $100$ and $1000$) lead to decreased stability in \name, as highlighted in Section~\ref{sec:utility_function_section}. Furthermore, we found that only with $D=2$ we could satisfy all the requirements that are feasible to satisfy, even under extreme high congestion ($45$ Mbps capacity). 

%I am here

\subsection{Fairness with Other Protocols}
We evaluate \name connections in comparison to other CC protocols: CUBIC, BBR, Reno and Vivace, sharing a $95$ Mbps bottleneck. This evaluation includes two \name connections, one requiring $100$ Mbps and the other $5$ Mbps, alongside a single connection requiring 100 Mbps using one of the aforementioned CC protocols.

\begin{figure}%[htbp]
 \centering
\vspace{-0.4cm}
\begin{subfigure}{0.45\linewidth}
  \centering
  \includegraphics[width=\linewidth]{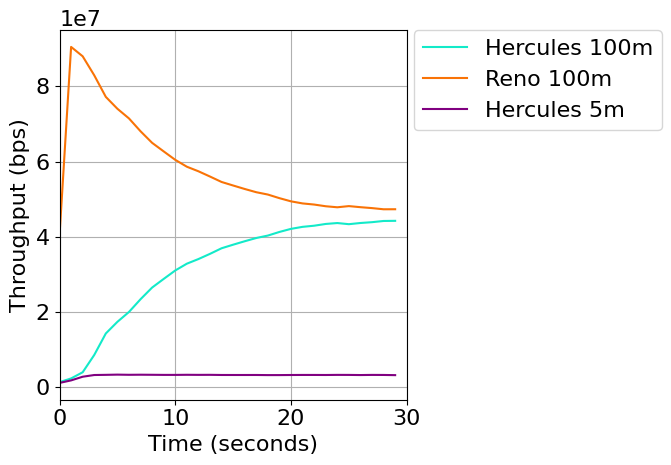}
  \vspace{-0.5cm}
  \caption{Reno}
  \label{fig:Hercules_Reno_scav}
\end{subfigure}%
\begin{subfigure}{0.45\linewidth}
  \centering
  \includegraphics[width=\linewidth]{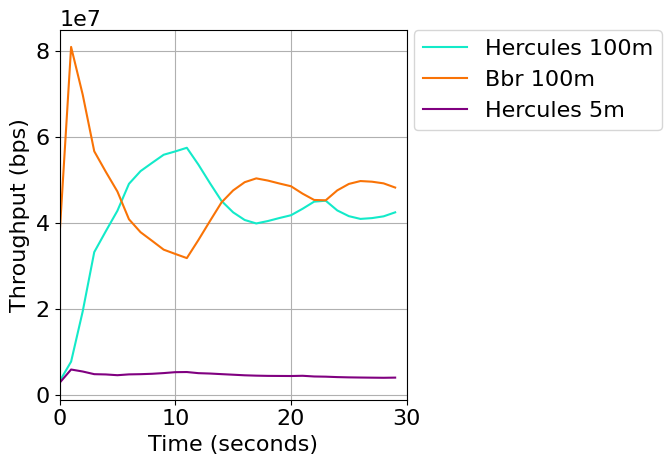}
  \vspace{-0.5cm}
  \caption{BBR}
  \label{fig:Hercules BBR and scav}
  %\vspace{-0.3cm}
\end{subfigure}%

%hspace{1em}
\begin{subfigure}{0.45\linewidth}
  \centering
  \includegraphics[width=\linewidth]{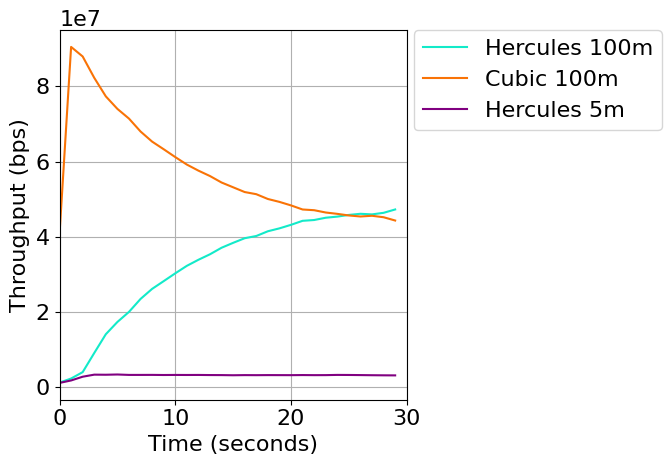}
  \vspace{-0.5cm}
  \caption{CUBIC}
  \label{fig:Hercules CUBIC and scav}
%\vspace{-0.3cm}
\end{subfigure}%
%\hspace{1em}
\begin{subfigure}{0.45\linewidth}
  \centering
  \includegraphics[width=\linewidth]{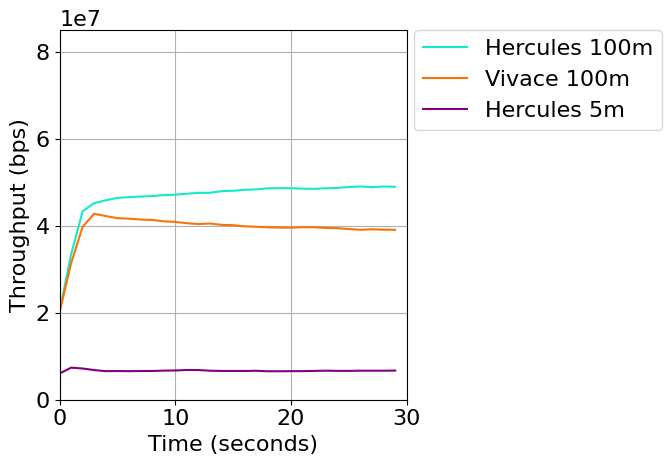}
  \vspace{-0.5cm}
  \caption{Vivace}
  \label{fig:Hercules Vivace and scav}
  %\vspace{-0.3cm}
\end{subfigure}%
%\vspace{-0.4cm}
\caption{Throughput convergence of Hercules alongside other CC protocols in a $95$ Mbps bottleneck.}
\label{fig:Hercules convergance with others}
\vspace{-0.3cm}
\end{figure}

%First, a static case is considered, where all the connections arrive to the network simultaneously. 
Fig.\ref{fig:Hercules convergance with others} illustrates the sending rates over time for \name connections (cyan and purple lines) and a connection using one of the mentioned CC protocol (orange line). 
%We can see that connections with same requirements, 
The graphs demonstrates that connections with the same bandwidth requirements converge to similar sending rates, regardless of the protocol used (whether \name or another). 
%Moreover, it shows that \myfair{} is achieved and the lower requirement connection is satisfied.
These findings remain consistent even in dynamic scenarios with non-simultaneous connections.
%Next, we considered a dynamic scenario where the \name low rate connection arrives $10$ seconds later. \neta{The results in} Fig.~\ref{fig:Hercules convergance with others dynamic} shows that the final convergence outcome and duration of all connections is similar to the static case.

%(Fig.~\ref{fig:Hercules CUBIC and scav}) and Vivace (Fig.~\ref{fig:Hercules Vivace and scav}) takes around $20-30$ seconds. Since BBR and Hercules are RTT sensitive, their convergence is slower. 
%Also, the fairness between Hercules, Vivace, CUBIC and BBR is demonstrated, as the connections which share the same requirements enjoy from the same sending rate.

\ignore{------------------------------------------------
\begin{figure}%[htbp]
 \centering
\vspace{-0.4cm}
\begin{subfigure}{0.3\linewidth}
  \centering
  \includegraphics[width=\linewidth]{figs/AWS_unlimited 100 5.0_boxplot_utilization.pdf}
  \vspace{-0.5cm}
  \caption{$95$ Mbps}
  \label{fig:Lab 100 5.0_boxplot_utilization}
\end{subfigure}%
%\hspace{1em}
\begin{subfigure}{0.3\linewidth}
  \centering
  \includegraphics[width=\linewidth]{figs/AWS_unlimited 125 5.0_boxplot_utilization.pdf}
  \vspace{-0.5cm}
  \caption{$120$ Mbps}
  \label{fig:Lab 120 5.0_boxplot_utilization}
  %\vspace{-0.3cm}
\end{subfigure}%
%\hspace{1em}
\begin{subfigure}{0.3\linewidth}
  \centering
  \includegraphics[width=\linewidth]{figs/AWS_unlimited 140 5.0_boxplot_utilization.pdf}
  \vspace{-0.5cm}
  \caption{$135$ Mbps}
  \label{fig:Lab 140 5.0_boxplot_utilization}
  %\vspace{-0.3cm}
\end{subfigure}%
%\vspace{-0.4cm}
\caption{Averaged network utilization and standard deviation for each protocol under different congestion conditions in the unbounded connections scenario.}
\label{fig:Hercules Utilization}
%\vspace{-0.3cm}
\end{figure}
------------------------------------------------}

\section{Related work}\label{sec:related_work}
Traditional CC protocols on the Internet focus on fair-share among all senders \cite{vivace, BBR, cubic, COPA}.
Recently, there has been a growing need to address heterogeneous requirements  of senders, leading to the development of new CC protocols.  Many of these protocols consider only two connection classes: primary and background traffic. 
For primary connections, TCP CUBIC \cite{cubic} is often recommended, while background connections may use Low Extra Delay Background Transport (LEDBAT) \cite{ledbat_plus}. However, this approach requires prioritization decisions upfront before entering the network.
%TCP CUBIC \cite{cubic} was recommended for primary connections while Low Extra Delay Background Transport (LEDBAT) \cite{ledbat_plus} was used by the background connections \cite{Cubic_Ledbat_Overview}. However, according to this approach, the connections should be prioritized in advanced, before entering the network. 

A more flexible approach prioritizes connections based on a threshold \cite{proteus}, allowing them to dynamically switch between  two prioritization classes over time.
Another strategy suggested running four distinct CC protocols simultaneously to create four priority levels or requirement levels \cite{PRISM}.
However, this approach is tailored for specific use-case of video streaming which includes  
%where multiple connections of a single connection require 
specific requirement classes.

Generalized CC protocols inspired by economic models have also emerged, aiming for a more comprehensive approach \cite{CC_shares,FairCloud_CC}.  In these protocols, each connection is allocated a ''share`` of the bandwidth. During congestion, the bandwidth allocation is based on the relative shares of each connection. However, this method doesn't directly consider application-specific requirements in its allocation process and requires perfect knowledge of the network.
%, which should reflect the economic arrangement that finances the Internet infrastructure. 
%However, this approach does not directly relate to application requirements.

\section{Conclusion and Further Research}\label{sec:conclusion}
In this work, we discussed and defined the desired fairness among  heterogeneous connections which we named \myfair{}.
Subsequently, we introduced a new and provable CC protocol named  ''\name``, designed to address heterogeneous requirements of senders in order to achieve an approximation of  \myfair{}. 

Our findings indicate that \name outperformed the state-of-the-art protocols by providing higher satisfaction ratio across all scenarios and network conditions. 
Furthermore, our evaluation highlights the fast convergence among \name' connections and 
adaptability to network changes as well as robustness to random loss and shallow buffers.  %\neta{Additionally, \name demonstrated fair treatment of other CC protocols, though further research could improve this fairness and reduce the time needed to achieve it.}
%\name interface receives connection requirements explicitly from the applications. A natural extension to this work can be to use new and existing ways for classifying connections and inferring requirements based on the traffic and the application which generates it.
Additionally, \name exhibits fairness also when operating along side other CC protocols, though further research could improve the time needed to achieve it.

Another research direction is to use \name exposure to traffic content and use it to dynamically infer connection requirements.

%for expanding this work is to enhance the \name interface to autonomously gather and infer connection requirements from traffic patterns and application characteristics. This could involve leveraging both established and novel methods for classifying connections based on their traffic attributes and inferring requirements dynamically.

\bibliographystyle{IEEEtran}
\bibliography{hercules_bib_short.bib}

\begin{thebibliography}{10}
\providecommand{\url}[1]{#1}
\csname url@samestyle\endcsname
\providecommand{\newblock}{\relax}
\providecommand{\bibinfo}[2]{#2}
\providecommand{\BIBentrySTDinterwordspacing}{\spaceskip=0pt\relax}
\providecommand{\BIBentryALTinterwordstretchfactor}{4}
\providecommand{\BIBentryALTinterwordspacing}{\spaceskip=\fontdimen2\font plus
\BIBentryALTinterwordstretchfactor\fontdimen3\font minus
  \fontdimen4\font\relax}
\providecommand{\BIBforeignlanguage}[2]{{%
\expandafter\ifx\csname l@#1\endcsname\relax
\typeout{** WARNING: IEEEtran.bst: No hyphenation pattern has been}%
\typeout{** loaded for the language `#1'. Using the pattern for}%
\typeout{** the default language instead.}%
\else
\language=\csname l@#1\endcsname
\fi
#2}}
\providecommand{\BIBdecl}{\relax}
\BIBdecl

\bibitem{6G_congestion_1}
M.~H. Alsharif, A.~Jahid, R.~Kannadasan, and M.-K. Kim, ``Unleashing the
  potential of sixth generation (6g) wireless networks in smart energy grid
  management: A comprehensive review,'' \emph{Energy Reports}, vol.~11, 2024.

\bibitem{6G_congestion_2}
I.~Akyildiz, A.~Kak, and S.~Nie, ``6g and beyond: The future of wireless
  communications systems,'' \emph{IEEE Access}, vol.~PP, pp. 1--1, 07 2020.

\bibitem{congestion_6G}
S.~Khan, A.~Hussain, S.~Nazir, F.~Khan, A.~Oad, and M.~D. Alshehri, ``Efficient
  and reliable hybrid deep learning-enabled model for congestion control in
  5g/6g networks,'' \emph{Computer Communications}, vol. 182, 2022.

\bibitem{IoT_CC1}
A.~P., H.~Vimala, and S.~J., ``Comprehensive review on congestion detection,
  alleviation, and control for iot networks,'' \emph{Journal of Network and
  Computer Applications}, vol. 221, p. 103749, 2024.

\bibitem{BW_vs_delay1}
E.~O'Connell, D.~Moore, and T.~Newe, ``Challenges associated with implementing
  5g in manufacturing,'' \emph{Telecom}, vol.~1, pp. 48--67, 06 2020.

\bibitem{BW_vs_delay_sensitive}
V.~Jordán, H.~Galperin, W.~Peres, and M.~Hilbert, ``Fast-tracking the digital
  revolution: Broadband for latin america and the caribbean,'' 2023.

\bibitem{BBR}
N.~Cardwell, Y.~Cheng, C.~S. Gunn, S.~H. Yeganeh, and V.~Jacobson, ``Bbr:
  Congestion-based congestion control,'' \emph{Queue}, vol.~14, no.~5, 2016.

\bibitem{cubic}
S.~Ha, I.~Rhee, and L.~Xu, ``Cubic: A new tcp-friendly high-speed tcp
  variant,'' \emph{SIGOPS Oper. Syst. Rev.}, vol.~42, no.~5, pp. 64--74, Jul.
  2008.

\bibitem{PCC}
M.~Dong, Q.~Li, D.~Zarchy, P.~B. Godfrey, and M.~Schapira, ``Pcc:
  Re-architecting congestion control for consistent high performance,'' in
  \emph{NSDI 15}, 2015, pp. 395--408.

\bibitem{vivace}
M.~Dong, T.~Meng, D.~Zarchy, E.~Arslan, Y.~Gilad, B.~Godfrey, and M.~Schapira,
  ``Vivace: Online-learning congestion control,'' in \emph{NSDI}, 2018.

\bibitem{COPA}
V.~Arun and H.~Balakrishnan, ``Copa: Practical delay-based congestion control
  for the internet,'' in \emph{NSDI 18)}, 2018, pp. 329--342.

\bibitem{CC_shares}
L.~Brown, G.~Ananthanarayanan, E.~Katz-Bassett, A.~Krishnamurthy, S.~Ratnasamy,
  M.~Schapira, and S.~Shenker, ``On the future of congestion control for the
  public internet,'' ser. HotNets '20.\hskip 1em plus 0.5em minus 0.4em\relax
  Association for Computing Machinery, 2020, p. 30–37.

\bibitem{FairCloud_CC}
L.~Popa, G.~Kumar, M.~Chowdhury, A.~Krishnamurthy, S.~Ratnasamy, and I.~Stoica,
  ``Faircloud: Sharing the network in cloud computing,'' \emph{SIGCOMM Comput.
  Commun. Rev.}, vol.~42, no.~4, p. 187–198, 2012.

\bibitem{Cubic_Ledbat_Overview}
M.~Geist and B.~Jaeger, ``Overview of tcp congestion control algorithms,'' in
  \emph{Network Architectures and Services}, May 2019.

\bibitem{ledbat_plus}
O.~N. Ertugay, D.~Havey, and P.~Balasubramanian, ``Ledbat++: Congestion control
  for background traffic,'' 2020.

\bibitem{proteus}
T.~Meng, N.~Rozen-Schiff, P.~B. Godfrey, and M.~Schapira, ``Pcc proteus:
  Scavenger transport and beyond,'' ser. SIGCOMM '20.\hskip 1em plus 0.5em
  minus 0.4em\relax Association for Computing Machinery, 2020, p. 615–631.

\bibitem{IoT_CC2}
T.~Kavitha, P.~Nagarajan, R.~Shobana, V.~Vinothini, S.~Karuppanan, A.~Jeyam,
  and A.~Malar, ``Data congestion control framework in wireless sensor network
  in iot enabled intelligent transportation system,'' \emph{Measurement:
  Sensors}, vol.~24, p. 100563, 11 2022.

\bibitem{IoT_CC3}
L.~P. Verma and M.~Kumar, ``An iot based congestion control algorithm,''
  \emph{Internet of Things}, vol.~9, p. 100157, 2020.

\bibitem{PRISM}
N.~Rozen-Schiff, A.~Navon, L.~Bruckman, and I.~Pechtalt, ``Prism based
  transport: How networks can boost qos for advanced video services?'' in
  \emph{Proceedings of the Workshop on Design, Deployment, and Evaluation of
  Network-Assisted Video Streaming}, ser. VisNEXT'21, 2021, p. 1–7.

\bibitem{Hercules_code}
\BIBentryALTinterwordspacing
I.~Pechtalt, N.~Rozen-Schiff, A.~Navon, and L.~Bruckman, ``{Hercules' code},''
  2024. [Online]. Available: \url{https://github.com/itzcak/Hercules}
\BIBentrySTDinterwordspacing

\bibitem{MPCC}
T.~Gilad, N.~Rozen-Schiff, P.~B. Godfrey, C.~Raiciu, and M.~Schapira, ``Mpcc:
  Online learning multipath transport,'' in \emph{CoNEXT}, 2020.

\bibitem{max_min2}
L.~Jose, S.~Ibanez, M.~Alizadeh, and N.~McKeown, ``A distributed algorithm to
  calculate max-min fair rates without per-flow state,'' \emph{Proc. ACM Meas.
  Anal. Comput. Syst.}, vol.~3, no.~2, jun 2019.

\bibitem{LMMF}
W.~Ogryczak, ``Lexicographic max-min optimization for efficient and fair
  bandwidth allocation,'' 2007.

\bibitem{LMMF_exists}
C.~Li, T.~Wan, J.~Han, and W.~Jiang, ``Towards distributed lexicographically
  fair resource allocation with an indivisible constraint,''
  \emph{Mathematics}, vol.~10, no.~3, 2022.

\bibitem{traffic_classification1}
N.~Menezes and F.~Mello, ``Flow feature-based network traffic classification
  using machine learning,'' \emph{Journal of Information Security and
  Cryptography (Enigma)}, vol.~8, pp. 12--16, 12 2021.

\bibitem{traffic_classification2}
S.~U. Jafri, S.~Rao, V.~Shrivastav, and M.~Tawarmalani, ``Leo: Online
  {ML-based} traffic classification at {Multi-Terabit} line rate,'' in
  \emph{21st USENIX Symposium on Networked Systems Design and Implementation
  (NSDI 24)}.\hskip 1em plus 0.5em minus 0.4em\relax Santa Clara, CA: USENIX
  Association, Apr. 2024, pp. 1573--1591.

\bibitem{arctan_neu_networks}
J.~Lederer, ``Activation functions in artificial neural networks: A systematic
  overview,'' \emph{ArXiv}, vol. abs/2101.09957, 2021.

\bibitem{Even_dar}
E.~Even-dar, Y.~Mansour, and U.~Nadav, ``On the convergence of regret
  minimization dynamics in concave games,'' in \emph{STOC '09}, 2009.

\end{thebibliography}
\end{document}